# Why Feedback Literacy Matters for Learning Analytics

Yi-Shan Tsai, Monash University, yi-shan.tsai@monash.edu

**Abstract:** Learning analytics (LA) provides data-driven feedback that aims to improve learning and inform action. For learners, LA-based feedback may scaffold self-regulated learning skills, which are crucial to learning success. For teachers, LA-based feedback may help the evaluation of teaching effects and the need for interventions. However, the current development of LA has presented problems related to the cognitive, social-affective, and structural dimensions of feedback. In light of this, this position paper argues that attention needs to shift from the design of LA as a feedback product to one that facilitates a process in which both teachers and students play active roles in meaning-making. To this end, implications for feedback literacy in the context of LA are discussed.

## Introduction

Learning analytics (LA) emerged as an interdisciplinary field a decade ago with a key mission to advance the understanding of learning and enhance this by using large datasets generated by learning processes. Early LA examples were largely motivated by the need to improve student retention in higher education, with a particularly common approach the application of predictive modelling to identify risks of failing (e.g., Arnold & Pistilli, 2012). Over the years, there has been growing awareness of the importance to ground LA in educational theories to identify meaningful metrics for the measurement of a learning phenomenon (Gašević et al., 2017). However, approaches to LA have been largely driven by behaviourism, in part due to constraints related to sources of data available; namely, commonly collectable data tend to focus on learner behaviour, such as log-ins and clicks. Despite this constraint, recent work in LA has made substantial break-throughs in inferring cognitive and meta-cognitive development among learners using LA (e.g., Jovanović et al., 2017; Uzir et al., 2020). However, the adoption of LA in real-world settings remains small in scale, thus limiting its expected impact on learning (Tsai et al., 2020; Viberg et al., 2018). Although there is abundant literature exploring social-technical issues that impede LA adoption (e.g., Macfadyen et al., 2014; Prinsloo & Slade, 2017; Tsai et al., 2021), there has been relatively little attention to key skills required for teachers and students to benefit from LA as a feedback process. A recent study by Gray and others (2021) proposed a framework for continuous professional development in LA for students, professional services, and teaching staff, focusing on three phases of skill development, including using and practicing feedback. However, the study has not explored skills that may be required for teachers and students to make sense of and use feedback generated by LA. This position paper is intended to bridge the gap by conceptualising feedback literacy in the context of LA for the purpose of enhancing the effectiveness and sustainability of LA and, more importantly, improving the overall feedback processes in higher education. To this end, I first conceptualise feedback and feedback literacy before outlining issues with LA-based feedback. Following that, I argue the importance of installing LA-specific feedback literacy among teachers and learners by highlighting three key dimensions required to make LA feedback effective.

## Conceptualising feedback and feedback literacy

Feedback plays an important role in learning, and the main purpose is to bridge the gap between a desired learning goal and the current performance (Evans, 2013). According to Hattie and Timperley (2007), feedback serves to answer three broad questions: Where am I going (*feeding up* to learning goals or expected standards that a learner works towards)? How am I going (*feeding back* on a learner's actual level performance in relation to goals and standards)? Where to next (*feeding forward to* greater possibilities for learning)? Traditionally, feedback is perceived as *information* generated and delivered from teachers to students. This transmission model is based on a cognitive perspective, and teachers are considered the ones mainly responsible for the effectiveness of feedback (Boud & Molloy, 2013; Evans, 2013). This conceptualisation is still very much present in ways teaching quality and student experience are measured in higher education (Winstone & Carless, 2021), which to some extent have also influenced how teachers and students see feedback (Winstone et al., 2021). While not mutually exclusive, compared to the cognitive perspective, the social-constructivist paradigm highlights the importance of feedback as a *process* that requires teachers and students to work in tandem, i.e., to have shared responsibilities (Boud & Molloy, 2013; Evans, 2013). For example, Yang and Carless (2013) discussed three inter-related dimensions of feedback: *cognitive, social-affective, and structural*. At the *cognitive* level, feedback needs to help students narrow the gap between their current and desired performance in addition to increase their self-regulated learning (SRL) skills. However, activities that happen in the cognitive dimension can be subject to the *social-affective* dimension;

that is, the power relationships between agents of feedback (e.g., teachers, students, and peers) and emotional reactions to feedback. As argued by Price et al. (2010, p. 280), "if the feedback is viewed as a product rather than part of a relational process, it is less likely to generate a response". In other words, a trust relationship between teachers and students is considered crucial to productive engagement with feedback, and treating feedback as a dialogic process may help offset power imbalance and cultivate trust (Yang & Carless, 2013).

The *structural* dimension is related to how feedback is organised and managed by the teacher and institution, e.g., timing, frequency, sequence, and modes of feedback. This dimension situates feedback in a wider learning context in which resources and policies can be determining factors of feedback effectiveness. When evaluations of feedback quality overemphasise this dimension, sustainability can become an issue because institutional resources are finite. In light of this, Boud and Molloy (2013) argue that for feedback to be sustainable, students need to be seen as active agents of change rather than passive recipients. Similarly, Winstone et al. (2021) suggest that a culture of responsibility sharing in feedback processes needs to be developed in order to move from feedback as telling to feedback as dialogue. On the other hand, Carless and Winstone (2020) suggest that in addition to *design* and *relational* dimensions, teacher feedback literacy has a *pragmatic* dimension; that is, a teacher's ability to be able to navigate tensions resulting from practical issues, such as resource and time constraints, by making inevitable compromises between the ideal, the defensible, and desirable feedback or leveraging technology for efficiency and timeliness.

Although both teacher-centred feedback and student-centred feedback highlight the importance of action in closing a feedback loop, the former assumes that learners depend on others to drive their learning and that learners would know what action to take when being provided with diagnostic information about their learning. In contrast, the latter emphasises the agency of learners who facilitate their own learning through active information seeking and sense-making (Boud & Molloy, 2013). Drawing on this student-centred view of feedback, Carless (2019) describes that feedback can take forms of *single, double*, and *spiral* loops. While feedback as a *single* loop focuses on short-term improvement of performance (predominantly task-driven), the *double*-loop process focuses on long-term impacts of feedback on the meta-cognition process, e.g., SRL, and the *spiral* process depicts learning as a complex, ongoing process in which learners continue to tackle some unresolved learning puzzles through continuous sense-making and the use of past feedback. In the core of this argument is the active role of learners who need to be able to *appreciate feedback*, *make judgement* based on feedback, *manage affect* resulted from feedback, and *take action* based on feedback – these are defined as key aspects of feedback literacy by Carless and Boud (2018). Building on this work, Molloy et al. (2020) further identified seven features of feedback literacy that emphasises feedback as an active, reciprocal process in which learners elicit information and enact outcomes of information processing, whereas Ryan et al (2021) proposed a model of eight feedback components that serve three functions: supporting future impact, enabling sensemaking, and supporting agency. In short, student-centred feedback processes requires feedback literacy to be present in both teachers and students, as have argued by Boud and Molloy (2013) that sustainable feedback depends on what learners bring and what the curriculum promotes in addition to what the environment affords. In this paper, feedback literacy for teachers is seen as *the ability to design and implement feedback not only as a piece of information that can help students achieve desired learning goals and personal development, but also as a process that serves pedagogical, social, and political purposes*; whereas feedback literacy for students is perceived as *the ability to seek feedback, decode and encode meanings of feedback relevant to their learning (thus resulting in beliefs and knowledge updates or action), and engage in feedback as a socio-cultural process*.

## Learning analytics-based feedback

Learning analytics (LA) is commonly defined as "the measurement, collection, analysis and reporting of data about learners and their contexts, for purposes of understanding and optimizing learning and the environments in which it occurs" (Long et al., 2011, p. 3). A LA cycle is described with four key elements: learners, data, metrics, and interventions (Clow, 2012). As learners interact with learning activities and environments, a range of data such as clickstream, login, forum discussion, academic performance, and sensory data, can be processed into metrics or analytics, often visualised in forms of dashboards as feedback for teachers or learners. The intention is to provide information useful to inform interventions (e.g., curriculum updates) or conversation with students (Bennett & Folley, 2019; De Laet et al., 2020; Wise, 2014). In line with Winne and Butler (1994) who argue that feedback is an inherent catalyst of SRL, LA-based feedback serves to trigger an iterative cycle of self-monitoring (internal feedback), performance, evaluation, and external feedback. For learners, LA as external feedback may prompt them to assess their knowledge and beliefs, including domain knowledge, strategy knowledge, and learning motivations, based on which learners may devise learning plans. Similarly, for teachers, LA-based feedback may prompt them to assess their knowledge and beliefs about their domain knowledge, pedagogical knowledge and strategy, and their current understanding of the learners, thereby reflecting on the effectiveness of teaching approaches and materials and the need to make change or provide interventions.

Unlike traditional feedback that typically involves teachers and learners, LA-based feedback includes the *algorithm* as a key agent that takes part in the process of measuring and evaluating learning outputs and communicating the results and feedback (Pardo, 2018). If feedback is directed to learners (student-facing LA), learners are expected to be able to make sense of the feedback and devise actions based on the presented data. In other words, there is an implied assumption that users have a functional level of feedback literacy and data literacy. If LA-based feedback is directed to teachers (teacher-facing LA), teachers act as a mediator between LA and students. In other words, teachers will be responsible for making sense of LA-based feedback and deciding whether and how to provide further feedback to learners as well as ways to incorporate LA into teaching practice. In this process, the direct interaction learners have is with the teacher. Thus, LA can facilitate two different loops of feedback, each highlighting the roles of learners and teachers differently in the sense-making process of LA-based feedback (Figure 1). In both loops, the outputs from learners (e.g., behaviour, performance, and emotional states after feedback is received) continue to feed back to LA, thus forming continuous feedback processes.

**Figure 1**
Two loops of learning analytics (LA) feedback directed at teachers (T) and learners (L) respectively

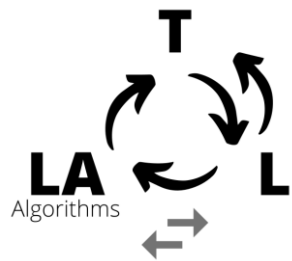

## Problems with learning analytics-based feedback

Although students generally welcome timely feedback enabled by LA to keep them on track of their learning (Galaige et al., 2018; Tsai, Mello, et al., 2021), current practice has presented some issues threatening the effectiveness of feedback. In the following, I frame these issues within the feedback triangle proposed by Yang & Carless (2013): *cognitive, social-affective*, and *structural* dimensions.

Considering the *cognitive* dimension, LA appears to focus more on *feeding back* (learning progress) and *feeding up* (learning goals), but less on *feeding forwards* (where to next). For example, a systematic review by Matcha et al. (2019) on 29 LA dashboards found that 18 focus on presenting learning progress and the most common indicators are numbers of logins, posts, and tweets, followed by indicators of domain knowledge, e.g., exam marks. Similarly, a review on 95 articles about student-facing analytics system (Bodily & Verbert, 2017) found that when text feedback was provided, it tended to be descriptive about what happened in the past and how students were doing up to the point of the feedback. As the most referenced theoretical concepts of learning analytics dashboards (LADs) design are SRL, reference frames are common techniques adopted to raise awareness and prompt reflections on one's learning progress, such as through social comparison with peers, comparison between current achievement and goals (the *feeding up* element), and comparison with the earlier self (Jivet et al., 2017). Compared to *feeding back* and *feeding up*, the *feeding forward* element has been an area of weakness in LA as empirical studies continue to report issues about lacking actionable insights (Bennett & Folley, 2019; Jivet et al., 2020; Lim et al., 2019). For example, Corrin and de Barba (2015) found that students struggled to translate LA into learning strategies; Jivet et al. (2017) identified little support of LADs in scaffolding goal setting and planning; Cha and Park (2019) reported that students desire prescriptive tips and suggestions to help them achieve desired goals; and Li et al. (2021) found that teachers struggled to infer student learning or identify learning problems from analytics results. As such, aligning LA with pedagogical practices and learning sciences continues to be a focused area of development in the field (Knight et al., 2014, 2020; Wise, 2014).

The *social-affective* dimension of feedback emphasises the dialogic (Sutton, 2009) and relational (Price et al., 2010) nature of a feedback process, which requires trust and balanced power relationship between the provider and the recipient of feedback (Winstone et al., 2021). Although scholars have argued the value of LA in facilitating dialogue between teachers and students (De Laet et al., 2020; Wise, 2014), LA as a feedback process involves not only teachers and students, but also algorithms as a key feedback agent (Figure 1). As such, distrust in LA algorithms and data or facilitators of LA can hamper the dialogic and relational aspects of feedback. For example, Jones et al. (2020) argue that issues around data privacy can violate a trust relationship between students and institutions; whereas Tsai et al. (2021) outline three areas of trust issues related to LA from perspectives of teachers and students: 1) numbers are subjective, 2) fear of power diminution, and 3) design and implementation. The first area includes issues about data accuracy, bias, and incomplete representation of learning. The second area includes negative impacts on professional and learner agency, unfair judgement on learning or teaching practice, the inability to remain in control of one's data. The third area is related to practical issues, including workload, the interpretability and usefulness of LA, equity of treatment, and demotivation to learn. Similarly, other studies that look into the social-affective dimension of feedback based on LA have identified students' fear of being treated as numbers (Roberts et al., 2016; Tsai, Mello, et al., 2021), and Lewis et al (2021) pointed out further that students who felt being noticed as individuals were more willing to seek support and engage with learning when LA is used.

The *structural* dimension of feedback focuses on how feedback is integrated into the instructional design and institutional processes. The rise of LA is in part a response to the growing pressure on higher education institutions to demonstrate quality (Ferguson, 2012; Tsai et al., 2020) and LA has demonstrated promising results in providing feedback at scale in a timely manner (Pardo et al., 2019). However, adoption and impacts of LA have been limited (Ferguson & Clow, 2017; Viberg et al., 2018) partly due to structural issues such as the complexity of educational systems (Macfadyen et al., 2014; Tsai et al., 2019) and the emphasis on analytics over learning (Guzmán-Valenzuela et al., 2021; Williamson et al., 2020), in addition to some of the trust issues mentioned above. A related issue to higher education's mounting pressure on boosting student satisfaction with timely feedback is that LA has been predominantly perceived as a technological tool to support this agenda rather than a process that requires proper integration into pedagogical practice (Wise, 2014). As a result, not only does LA-based feedback tend to be perceived as a product in line with the traditional feedback model – feedback as telling, but evaluations of LA also often focus on assessing its usability (e.g., acceptance, usefulness and ease-of-use) and impact on behavioural change to the extent that it is observable and the data is capturable (Jivet et al., 2018). This dominant conceptualisation of LA as feedback overlooks the cognitive and social-affective dimensions of feedback. It may also be argued that LA continues to promote a transmission model of feedback (Boud & Molloy, 2013), which considers a feedback loop closed in so far as action is taken by learners regardless of whether it is a passive response without learning actually taking place.

## Learning analytics feedback literacy

Returning to the definition of LA as a process that aims to not only *understand*, but also *optimise* learning (Long et al., 2011), we need to move beyond evaluation to enhancement of learning. For example, before seeking better approaches to predict or measure learning results, ask first what learning gain can be expected by applying LA in teaching (Kitto et al., 2018). Before making algorithmic choices, ask first how the decisions align with human values (Chen & Zhu, 2019) and educational values (Wise et al., 2021). Before choosing what data to collect, ask first what metrics are meaningful to a teaching context (Brown, 2020). Before asking what students prefer for an LAD, ask what they need (Jivet et al., 2020). Important to all these questions is a thoughtful consideration of what teachers and learners offer. It is thus necessary to ask what feedback literacy means in a LA-based feedback process. In the following, I build on Carless and Boud (2018) and highlight three broad dimensions of feedback literacy that is important to effective feedback and the exertion of agency by both teachers and learners in the context of LA.

First, *appreciating LA-based feedback with critical awareness of its value and limitations*. To elicit a response from learners to feedback, it is crucial that learners recognise the value of feedback in learning so as to be receptive and proactive (Carless & Boud, 2018; Winstone et al., 2017). Similarly, the uptake of LA as data-driven feedback requires appreciation of data as computational representation of learners and learning, which can navigate attention to areas that may need intervention from the teacher or learners themselves (Gibson & Martinez-Maldonado, 2017; van Leeuwen et al., 2017). However, it is important to recognise that LA is valuable in so far as providing a snapshot of learning, the meaning of which is not complete until learners who generate the data and the teacher who facilitates learning activities bring in their interpretations drawing on their understanding of and experience in the learning or teaching situation. The knowledge of learners and teachers is particularly valuable in filling gaps of information that is not capturable by LA.

Second, *translating the computational representation of learning to psychological self or others and enacting the derived meanings*. As described earlier, LA-based feedback provides computational representations that describe what learners do or how they perform. Gibson and Martinez-Maldonado (2017) describe the sense-making process as an act of translation between the *computational epistemic domain* (computer) and the *psychosocial epistemic domain* (human). The former involves encoding what is observed from learners and processing it into a form (the computational representation) suitable for human interpretation. The latter involves a cognitive cycle of interpreting the computational representation in relation to the psychological *self* or *others*, thereby constructing meanings and turning this into further products, such as action or the creation of an artefact or digital traces. Thus, a key dimension of feedback literacy in an LA-feedback process is the ability to move between the two domains. This requires both teachers and learners to actively interpret analytics in relation to their situations (e.g., learning and teaching approaches) and decide what action to take or not to take.

Third, *managing affect and negotiating power in the feedback loop*. Same as other forms of feedback, LA-based feedback can induce stress, anxiety, frustration, and other negative emotions especially when feedback continues to flag underperformance or when comparisons are made between individuals and peers (e.g., see Jivet et al. (2020) and Lim et al. (2019)). In addition, as data processing is not simply computational, but also inherently social and political, various decisions are made by a range of stakeholders (e.g., researchers, developers, managers, and service providers) involved in the overall LA cycle – from choosing data and algorithms to delivering downstream intervention. As such, both learners and teachers need to be proactive in questioning whose interests are being served when interacting with LA-based feedback. Moreover, depending on how data is collected, processed, shared, and presented, the process can cause discomfort when a sense of surveillance is experienced (Kwet & Prinsloo, 2020; Selwyn, 2020). It is thus necessary to engage in dialogue with decision makers to offset imbalanced power relationships and increase transparency, validity, and accountability. In other words, being LA literate is not only about being able to make sense of LA-based feedback productively, but also about having the ability to engage with LA in the wider social-political context, such as interrogating the ownership of one's own data or decisions made about or for an individual.

## Conclusion

There is no doubt that LA has opened a door to a multitude of opportunities to understanding learning patterns and how learners interact with learning environments and others therein. It also has great potential to deal with one of the biggest feedback challenges – traceability (Malecka et al., 2020) – by iterating feedback loops through continuous data collection as shown in Figure 1. However, the effectiveness and sustainability of LA-based feedback are questionable if it does not support learners and teachers to exercise their agency or scaffold the development of LA-specific feedback literacy, which is conceptualised with three broad dimensions in this article:

- Appreciating LA-based feedback with critical awareness of its value and limitations
- Translating the computational representation of learning to psychological self or others and enacting the derived meanings
- Managing affect and negotiating power in the feedback loop

An important implication for practice is the need to scale training for both teachers and students. For the former, the training should focus not only on strengthening skills in the three aspects, but also related pedagogical skills including ways to enhance a feedback process with LA as part of the learning design and ways to support students to develop LA-feedback literacy. For students, such training may be the most effective if incorporated into their first-year curriculum so as to maximise their opportunities to benefit from LA-based feedback, as also suggested by Molloy et al. (2020) regarding feedback literacy development in higher education. There are also implications for LA design. Firstly, the design should prompt *dialogue*, such as showing reflective or even provocative questions to encourage thinking around important issues related to educational and human values, or embed communication channels for users to raise questions and express opinions to relevant stakeholders in an institution. Secondly, the design should *foreground sense-making*, such as providing mechanisms to guide interpretations in relation to where data comes from, what it means, and how it relates to oneself. Thirdly, the design should *support possibilities for further learning* (feeding forward)(Hattie & Timperley, 2007), such as providing recommendations based on inputs from teachers. In this way, LA may move away from a transmission feedback model to one that is student-centred, relational and process driven, with shared responsibilities between teachers and learners in making feedback effective.